\documentclass[pre,twocolumn,superscriptaddress,floatfix,showpacs,showkeys]{revtex4}

\usepackage[latin2]{inputenc}
\usepackage{amsmath}
\usepackage{amssymb}
\usepackage{epsfig}
\input epsf

\newcommand{\be}{\begin{equation}}
\newcommand{\bel}[1]{\begin{equation}\label{#1}}
\newcommand{\ee}{\end{equation}}
\newcommand{\bea}{\begin{eqnarray}}
\newcommand{\ba}{\begin{array}}
\newcommand{\eea}{\end{eqnarray}}
\newcommand{\ea}{\end{array}}

\begin{document}
\draft


\title{ Characteristics of Vehicular Traffic Flow at a Roundabout }
\author{M. Ebrahim Fouladvand, Zeinab Sadjadi and M. Reza Shaebani }

\address{ Department of Physics, Zanjan University, P.O.
Box 313, Zanjan, Iran.}

\date{\today}

\begin{abstract}
We construct a stochastic cellular automata model for the
description of vehicular traffic at a roundabout designed at the
intersection of two perpendicular streets. The vehicular traffic
is controlled by a self-organized scheme in which traffic lights
are absent. This controlling method incorporates a yield-at-entry
strategy for the approaching vehicles to the circulating traffic
flow in the roundabout. Vehicular dynamics is simulated within the
framework of the probabilistic cellular automata and the delay
experienced by the traffic at each individual street is evaluated
for specified time intervals. We discuss the impact of the
geometrical properties of the roundabout on the total delay. We
compare our results with traffic-light signalisation schemes, and
obtain the critical traffic volume over which the intersection is
optimally controlled through traffic light signalisation schemes.
\end{abstract}


\maketitle

\section{Introduction}

Undoubtedly, traffic management is nowadays considered as one of
the basic ingredients of modern societies and large sums are
invested by governments in order to increase its efficiency. The
rapidly growing volume of vehicular traffic flow, limitations on
expanding the construction of new infrastructure, hazardous
environmental impact due to the emission of pollutants, together
with unfavourable delays suffered in congested traffic jams, are
among the basic features which necessitate the search for new
control, as well as optimization schemes, for vehicular traffic
flow. Inevitably, this task would not be significantly fulfilled
unless a comprehensive survey of vehicular dynamics, within a
mathematical framework, is developed. This has motivated
physicists to carry out extensive numerical, as well as analytical
research, in the discipline of {\it traffic flow theory}.\\
The statistical physicist´s contribution to the field has
accelerated since 1990's when computers provided the possibility
of simulating traffic flow through the discretization of space and
time. Ever since a vast number of results, both analytically and
empirically, has emerged in traffic discipline
\cite{css99,helbbook,kerner,book,tgf95,tgf97,tgf99}. In principle,
the flow of vehicles can be regarded as an interacting system of
particles. The techniques developed in statistical physics, which
to a very great extent, can be applied to investigate the basic
properties associated with the underlying dynamics of vehicular
flow. Broadly speaking, the traffic flow theory can be categorized
into two parts: {\it high way traffic} and {\it city traffic}, and
now there is vast literature in both domains. In this paper we
focus out attention on a particular aspect of city traffic, the
so-called {\it roundabout}. We try to present a numerical
investigation on controlling urban traffic via roundabouts.
Traditionally, the conflicting flows in urban areas were
controlled by signalised intersections. Ever since the
installation of first traffic light in New York in 1914, the
subject of urban traffic and its optimal control has been
intensively explored. Nowadays, the traffic in most of urban areas
are controlled by signalised intersections. Modern roundabouts
have quite recently come into play as alternatives to signalised
intersections, which tend to control the traffic flow more
optimally and in a safer manner. A roundabout is a form of
intersection design and control which accommodates traffic flow in
one direction around a central island, operating with yield
control at the entry points, and giving priority to vehicles
within the roundabout (circulation flow). Several characteristics
such as safety, deflection and turning movements, and construction
costs, to name but a few, distinguish a modern roundabout from the
more general form of a traffic circle.\\
The era of modern roundabouts began in the United Kingdom in the
1950's with the construction of the first "yield-at-entry"
roundabouts. In 1966, a nationwide yield-at-entry rule launched
the modern roundabout revolution. Yield-at-entry is the most
important operational element of a modern roundabout, but it is
not the only one. Deflection of the vehicle path and entry flare
are also important characteristics that distinguish the modern
roundabout from the non-conforming traffic circle, which does not
have these characteristics. The primary characteristics of the
modern roundabout reduce many of the safety hazards of traditional
intersections and non-conforming traffic circles.\\
The physical configuration of a modern roundabout, with a
deflected entry and yield-at-entry, forces a driver to reduce
speed during the approach, entry, and movement within the
roundabout. This is contrary to an intersection where many drivers
are encouraged by a green or yellow light to accelerate to get
across the intersection quickly in order to "beat the red light",
and contrary to old traffic circles where tangent approaches also
encourage, or at least allow, high-speed entries. It has always
been a subject of argument whether to control a intersection under
a signalised or non-signalised scheme via roundabouts. Apparently
in low-volume situations, non-signalised methods seem to show
better performance; whereas, in high-volume traffic, one has to
apply traffic light signalisation
\cite{book,robertson,porche,huberman}. The basic question which
arises is under what circumstances should one control an
intersection by signalised traffic lights? To address this
fundamental question, we try to explore and analyze some basic
characteristics of a typical roundabout, such as flow and delay,
in order to find a quantitative understanding. In what follows we
try to illustrate these fascinating aspects through computer
simulations.

\section{ Description of the Problem}

We now turn to discussing the simulation of traffic at a
roundabout. A roundabout is a form of intersection design and
control which accommodates traffic flow in one direction around a
central island and gives priority to vehicles within the
roundabout (circulation flow). Let us first discuss the basic
driving principles applied to roundabouts. In its most general
form, a roundabout connects four incoming, as well as four
outgoing flow directions. In principle, each incoming vehicle
approaching the roundabout can exit from each of four out-going
directions via making appropriate turning maneuvers around the
central island of the roundabout. The following figure illustrates
the situation.

\begin{figure}
\centering
\includegraphics[width=9.0truecm]{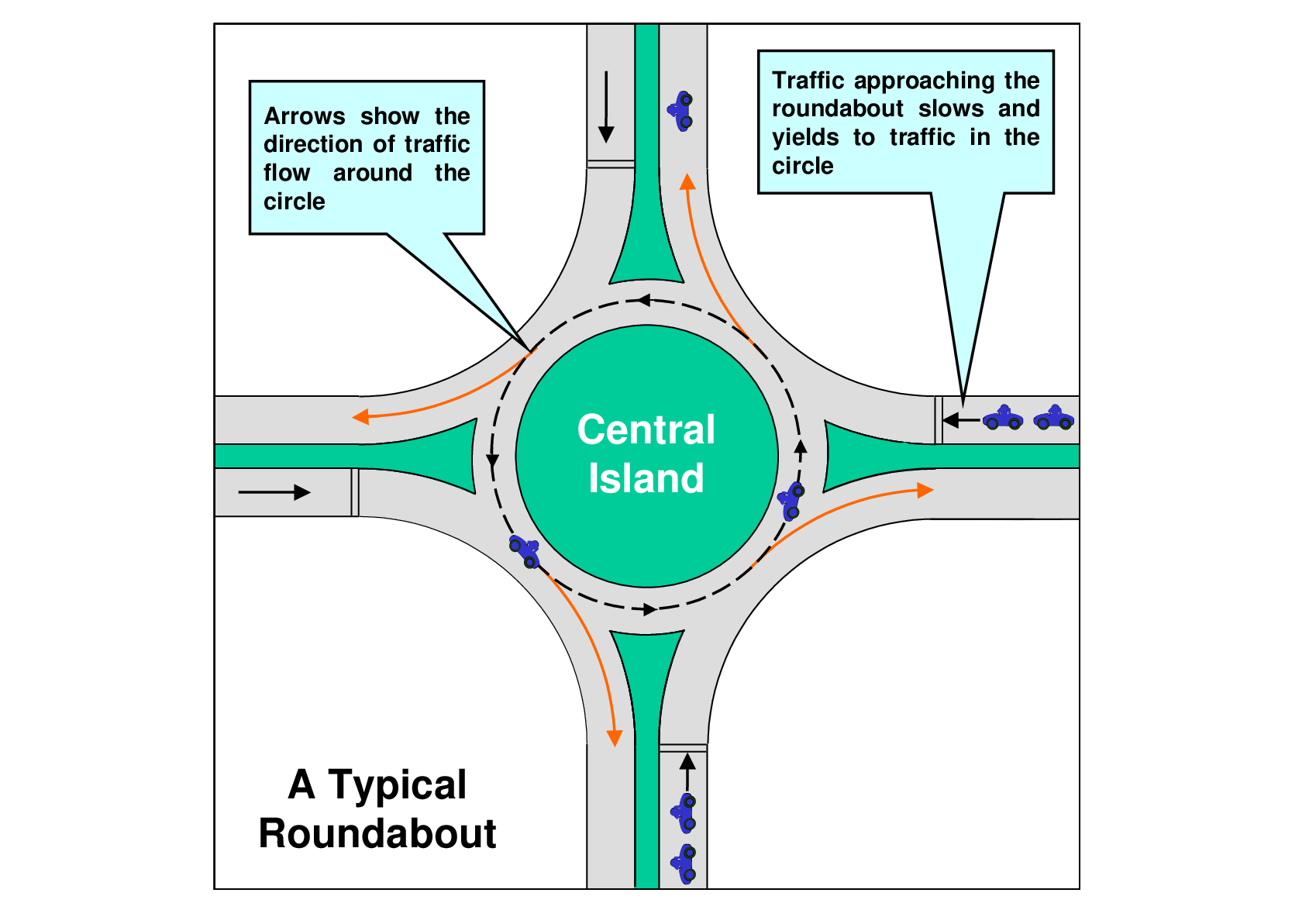}
\caption{A typical roundabout with yield-to-entry rules. Four 
exit directions are shown by arrows.}
\label{Fig1}
\end{figure}

The rules of the road give the movement priority to those vehicles
which are circulating around the central island. The approaching
vehicles should yield to the circulating traffic flow in the
roundabout, and are allowed to enter the roundabout, provided some
cautionary criteria are satisfied. In this paper we only
investigate a simplified version in which all the streets,
including the circulating lane around the central island, are
assumed to be single-lane. Let us explain the entrance regulations
in some details. Each approaching vehicle to the entry points of
the roundabout should decelerate and simultaneously look at the
left-ward quadrant of the roundabout. If there is any vehicle in
this quadrant, then the approaching vehicle should come to a
complete stop until the inside vehicle(s) leave(s) this quadrant.
The stopped car is only allowed to enter the roundabout provided
that no vehicle appears in its left side quadrant; otherwise, it
has to slow down and stop. The stopped direction can flow as soon
as the front car finds no car in the left-side quadrant of the
circulating lane.\\
This is possible due to stochastic fluctuation in the space gap of
the flowing direction. Once such an appropriate space gap has been
found, the stopped car is allowed to enter the roundabout. This
procedure is continuously applied to all approaching vehicles. Now
we return to those vehicles which are moving around the interior
island of the roundabout. Once a vehicle is permitted to enter the
roundabout, it continues moving until it reaches to its aimed exit
direction. Depending on the selected out-going direction, each
interior vehicle moves a portion of the way around the central
island. These turning movements are classified as: right-turn,
straight ahead, left-turn and U-turn. For those who tend to make a
U-turn, the whole circumference should be travelled. The interior
vehicle can freely move around the roundabout until it reaches the
desired exit. The above-mentioned driving rules establish a
mechanism responsible for controlling the traffic in conflicting
points. This mechanism blocks any direction which is conflicting
with a flowing one, thereby producing waiting queues in blocked
directions. In contrast to signalised intersections, intersecting
streets through roundabouts are controlled via a self-organized
mechanism of blocking.\\
It is evident that in congested traffic situations, where the
in-flow rates are high, the probability of finding a large space
gap is low. This leads to global blocking of other directions,
which in turn gives rise to the formation of pronounced queues. In
this situation, the roundabout performance is inefficient, and
apparently signalised control strategy shows a better performance.
Conversely, in relatively low traffic volume it is likely to find
a large space gap (by fluctuation) in the circulating direction
and hence, the possibility of entrance for the block direction
increases. This increases the efficiency of the roundabout. The
roundabout efficiency significantly depends on the incoming fluxes
of cars and statistics of space gaps. The basic question raised is
{\it under what circumstance the self-organized control scheme
becomes inefficient}? In order to find a better insight to the
problem, we have simulated the traffic flow and have investigated
the roundabout performance for various traffic situations and
geometrical sizes of roundabout. In the subsequent sections we
present our simulation results.

\section{ Formulation of the Model}

In this section we begin with the simplest flow-structure of the
roundabout. In this case, the roundabout connects two single-lane
one-way to one-way streets. With no loss of generality, we take
the direction of flows to be northward on street A and eastward on
street B (see the following figure). Also we give a permanent
priority to the flow of street B i.e., those cars which are
driving on street B can enter the roundabout without any caution.
On the contrary, the flow in street A should yield to the flow of
street B. A-vehicles should observe the yield-at-entry rules. They
are allowed to enter only by gap fluctuations in the B-flow.
Correspondingly, no delay is wasted for B-vehicles.

\begin{figure}\label{Fig2} \epsfxsize=11.5truecm
\centerline{\epsfbox{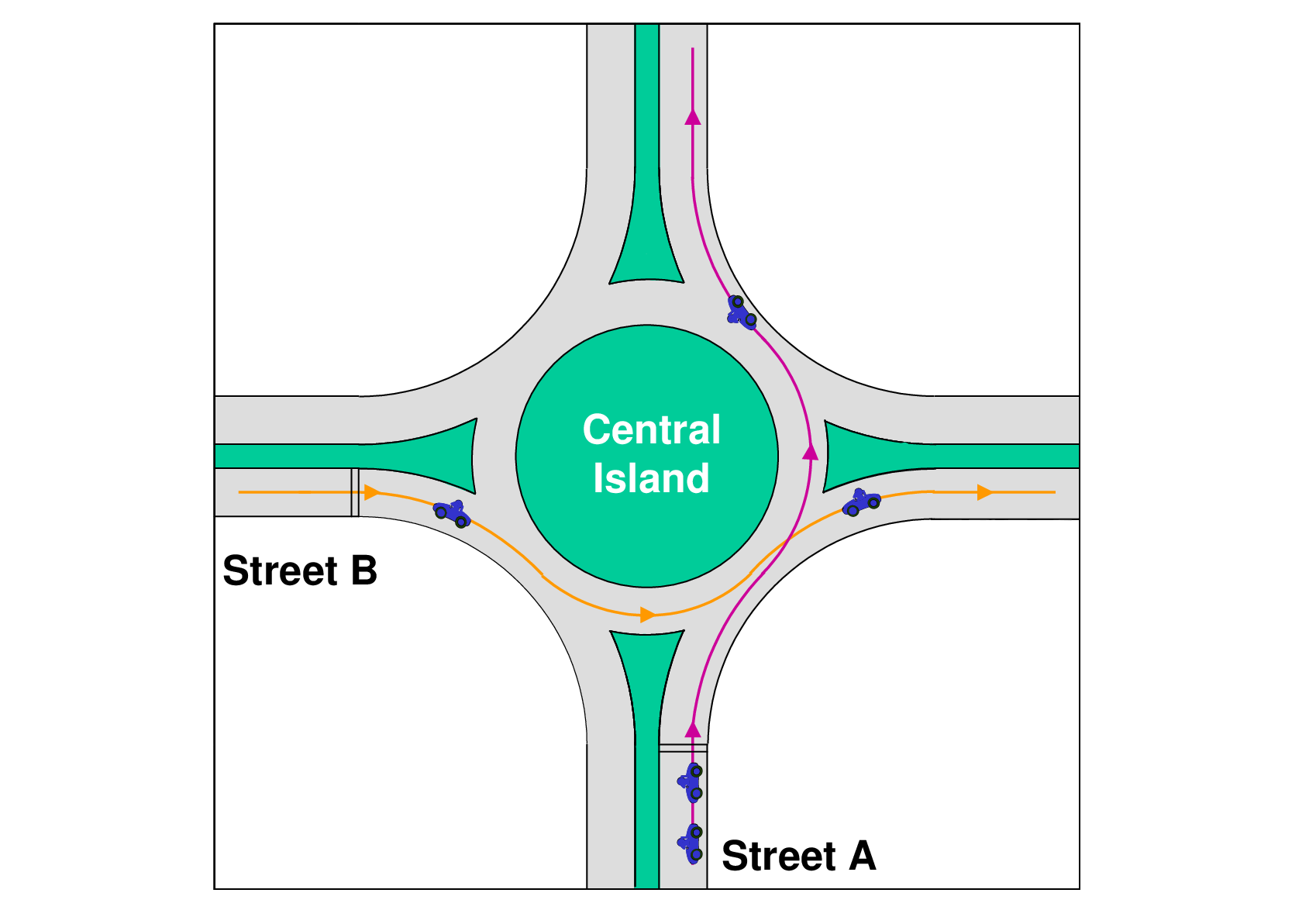}} \vspace{0.02cm} {\small{Fig.2:
simplest flow structure of a roundabout:
intersection of two one-way perpendicular streets. No turning is allowed. } }\\
\end{figure}

In order to capture the basic features of the problem, we have
simulated the traffic movement in the framework of cellular
automata. For this purpose, space and time are discretized in such
a way that each street is modelled by a one-dimensional chain
divided into cells which are the same size as a typical car
length. The circulating lane of the roundabout is also considered
as a discretized closed chain. Time is assumed to elapse in
discrete steps. We take the number of cells to be $L$ for both
streets and $L_r$ for the interior lane. Each cell can be either
occupied by a car or be empty. Moreover, each car can take
discrete-valued velocities $1,2,\cdots, v_{max}$. To be more
specific, at each step of time, the system is characterized by the
position and velocity configurations of the cars on each road. We
note that due to the turning maneuvers, the maximum velocity of
circulating cars should be reduced. Here we assume that the
maximum velocity for interior cars takes the value of 40 km/h.
The system evolves under a generalized discrete-time
Nagel-Schreckenberg (NS) dynamics. The generalized model
incorporates the anticipation effects of driving
habits\cite{knospe}. It modifies the standard NS model at its
second step i.e., adjusting the velocities according to the space
gap. Let us briefly explain the updating rules which are
synchronously applied to all the vehicles. We denote the position,
velocity and space gap (distance to its leading car) of a typical
car at discrete time t by $x^{(t)},v^{(t)}$ and $g^{(t)}$. The
same quantities for its leading car are denoted by
$x_l^{(t)},v_l^{(t)}$ and $g_l^{(t)}$. Assuming that the expected
velocity of the leading car, anticipated by the one following, in
the next time step $t+1$ takes the form
$v_{l,anti}^{(t)}=min(g_l^{(t)},v_l^{(t)})$, we define the
effective gap as $g_{eff}^{(t)}= g^{(t)} +
max(v_{l,anti}^{(t)}-gap_{secure},0)$ in which $gap_{secure}$ is
the minimal security gap. Concerning the above-mentioned
considerations, the following updating steps evolve the position
and the velocity of each car.\\

1) Acceleration:\\

$v^{(t+1/3)}= min(v^{(t)}+1, v_{max})$\\

2) Velocity adjustment :\\

$v^{(t+2/3)}=min(g_{eff}^{(t+1/3)}, v^{(t+1/3)})$\\

3) Random breaking with probability $p$:\\

if random $< p$ then $v^{(t+1)}=max(v^{(t+2/3)}-1,0)$\\

4) Movement : $x^{(t+1)} =x^{(t)}+ v^{(t+1)}$ \\

The state of the system at time $t+1$ is updated from that in time
$t$ by applying the modified NS dynamical rules. Let us now
specify the physical value of our time and space units. Ignoring
the possibility of existence of long vehicles such as buses,
trucks, etc., the length of each cell is taken to be 5.6 metres
which is the typical bumper-to-bumper distance of cars in a
waiting queue. Concerning the fact that in most of urban areas a
speed-limit of 60 kilometres/hour should  be kept by drivers, we
quantify the time step in such a way that $v_{max}=6$ corresponds
to the speed-limit value (taken as 60 km/h). In this regard, each
time step equals two seconds; and therefore, each discrete
increment of velocity signifies a value of 10km/h which is
equivalent to a comfortable acceleration of 1.4 $m/s^2$. We have
also set the streets lengths as $L=70$ cells and $gap_{sec}=1$. We
want to emphasize that the roundabout size i.e., the circumference
of the central island is a crucial parameter and should be
carefully treated. At the end of each updating step, we evaluate
the aggregate delay of street A. During the periods of the flow of
B-vehicles in the interior lane, the A-vehicles are hindered, and
accordingly, should stop before the interior island; hence, a
queue will be formed. As soon as a car comes to a halt, it
contributes to the total delay. In order to evaluate the delay, we
measure the queue length (the number of stopped cars) at time step
t, and denote it by the variable $Q$. Delay at time step $t+1$ is
obtained by adding the queue length $Q$ to the delay at time step
$t$. \be delay(t+1)= delay(t)+ Q(t) \ee

This ensures that during the next time-step, all of the stopped
cars contribute one step of time to the delay.

\subsection{ entrance of cars to the intersection}.

So far, we have dealt with those cars within the horizon of the
roundabout which goes up to the boundary points located at site
$L$ up-stream from each incoming flow. It would be illustrative to
discuss the entrance of cars into the roundabout. We take the
distance of the boundary position to be 70-cells, equivalent to
400 metres to the central island. The time head-ways between
entering cars at this entry location vary in a random manner which
consequently implies a random distance headway between successive
entering cars. As a candidate for describing the statistical
behaviour of the random space gap of entering cars, we have chosen
the Poisson distribution. The Poisson distribution function has
been used in a variety of phenomena incorporating the modelling of
"queue theories" and has proven to be a good estimation of reality
\cite{erlang}. In addition, it has the merit of taking only
discrete values which is desirable to us in the view of the fact
that in our model the gap is a discrete variable. According to
this distribution function the probability that the space gap
between the car entering the intersection horizon and its
predecessor be $n$ is : $ p(n) = \frac{\lambda^n e^{-\lambda}
}{n!} $ where the parameter $\lambda$ specifies the average as
well as the variance of distribution function. The parameter
$\lambda$ is a direct measurement of traffic volume. A large value
of $\lambda$ describes light traffic, while on the other hand, a
small-valued $\lambda$ corresponds to a heavy traffic state.

\subsection{ Simulation Results }

We let the roundabout evolve for 1800 time steps which is equal to
a real time period of one hour. We have averaged the results of 50
independent runs. Let us first consider the symmetric traffic
states in which the traffic conditions are equal for both roads.
In this case, we load the streets equally with approaching cars,
spatially separated by a random space gap (Poisson statistics)
from each other. Figure (3) depicts the total delay curves as a
function of average space gap of entering cars
$\lambda_A=\lambda_B=\lambda$ for various roundabout sizes. All
vehicles leave the roundabout along the incoming direction viz.
they are not permitted turn right, left or U-turn upon circulating
the roundabout. According to the graph, the delay shows a rapid
decline for light traffic states. This marks the high efficiency
of roundabout in low-volume traffic situations. Roundabouts are
designed in different sizes to serve various objectives and
conditions. Even mini-roundabouts are effective at reducing speed
and improving safety. Our simulation results confirm that
roundabout size plays a dominant role in its performance. Figure
three suggests the short-sized roundabouts operate more optimally.
We next examine the impact of asymmetry in the traffic volumes of
the streets. For this purpose, we fix the in-flow rate in street B
at $\lambda_B=13$ cells and vary the in-flow rate of street A.
Figure four depicts the behaviour of delay in terms of
$\lambda_A$.

\begin{figure}\label{Fig3}
\epsfxsize=7.6truecm
\centering
\epsfbox{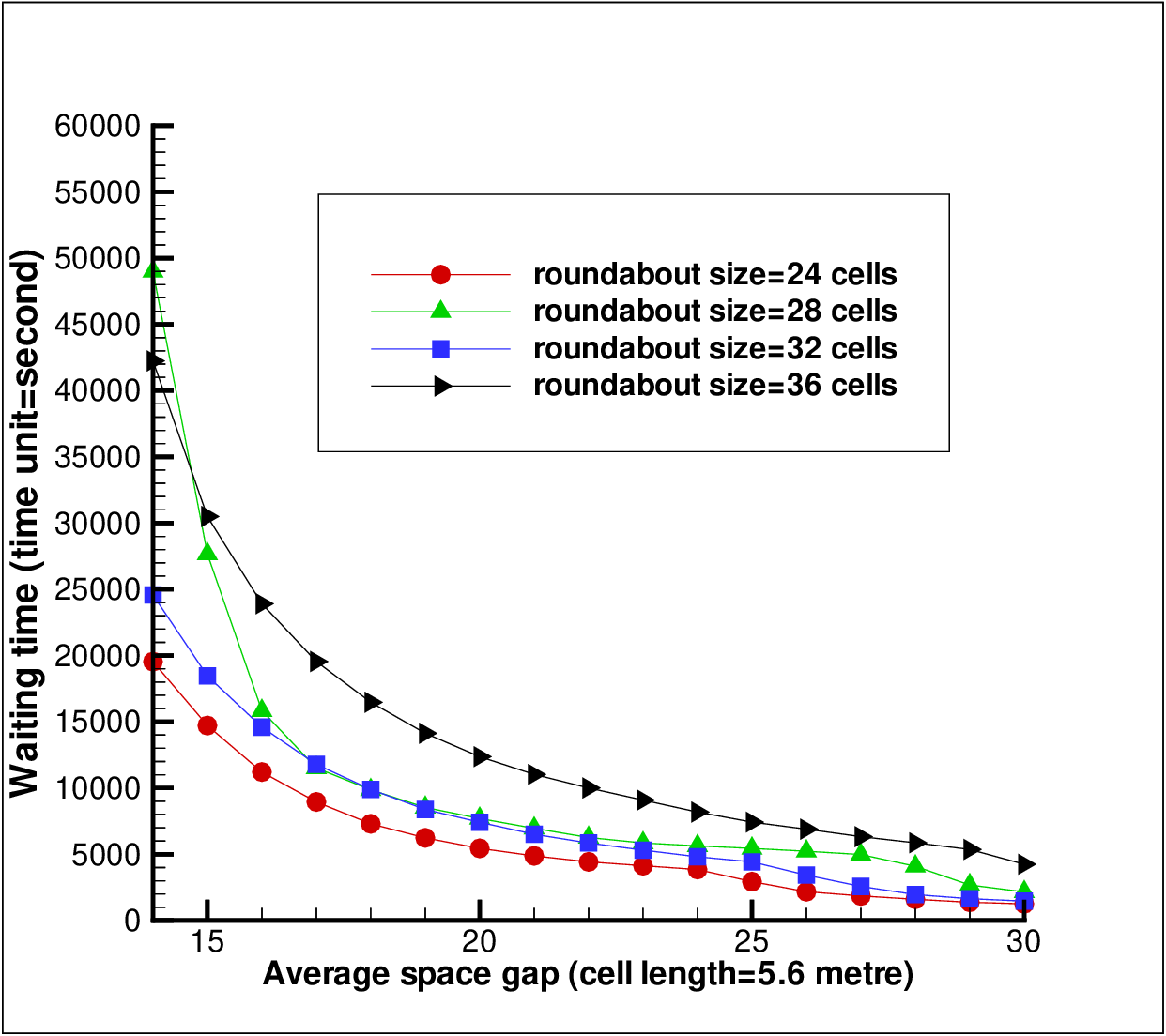}
\vspace{0.02cm}
\end{figure}

\begin{figure}\label{Fig4}
\epsfxsize=7.6truecm 
\centering
\epsfbox{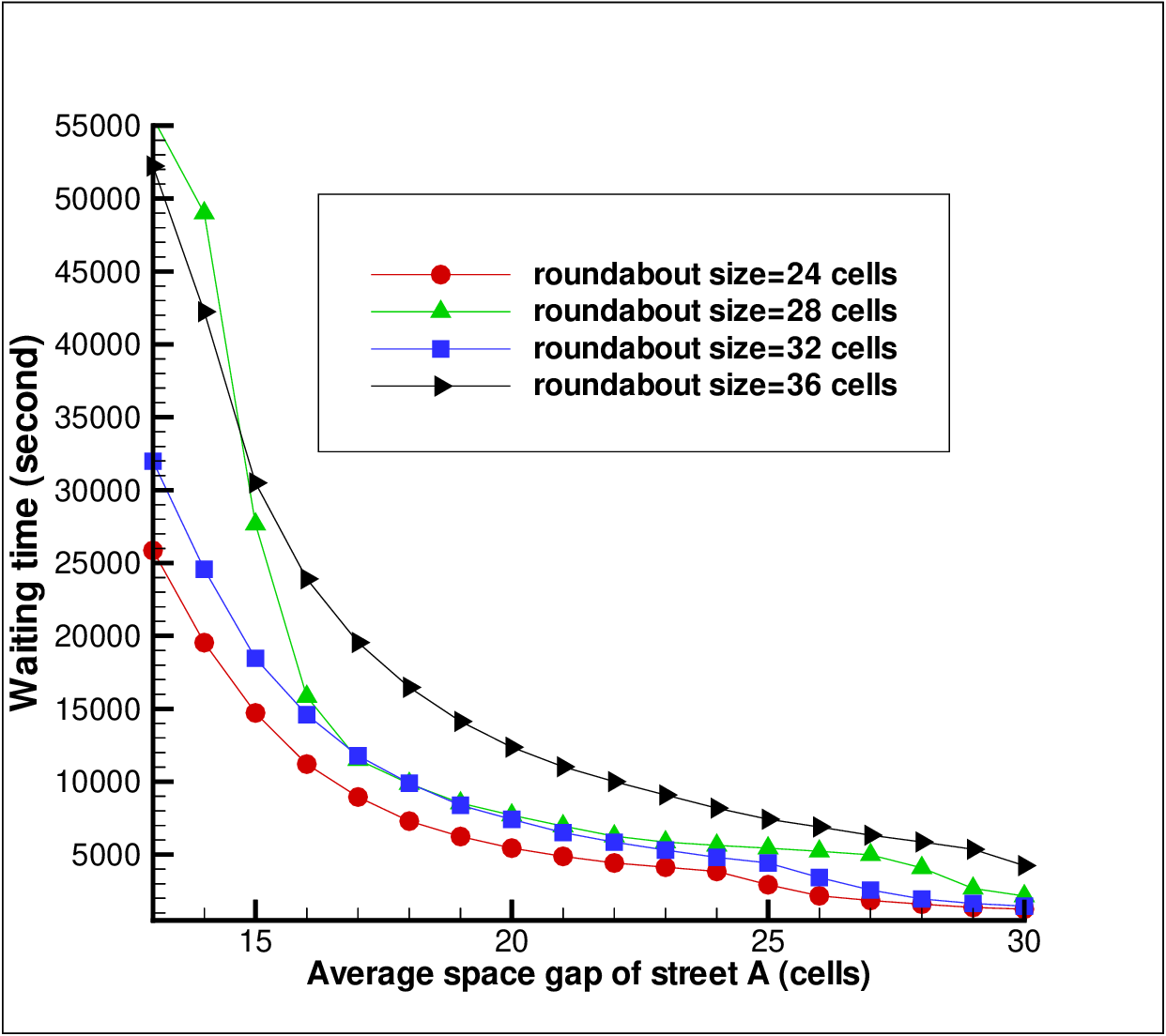}
\vspace{0.02cm} \small{Fig.~3,4: Total delay versus the average
space gap of approaching cars (top) and that of street A (bottom)
for various
sizes of roundabout. $\lambda_B=13$ in the bottom graph.} \\
\end{figure}

\section{ Right-Turn Permission }

At this stage, we remove parts of the restriction on the exit
direction and enable each car to leave the roundabout at its first
exit i.e., a right-turn. This implies that the south-north
direction (street A) is equipped with an extra south-bound lane
along which the incoming B-vehicle can leave the roundabout
through a right-turn. The following figure illustrates the
situation. Analogously, the approaching A-vehicles can leave the
roundabout through the exit leg of street B via a right-turn
maneuver. Therefore, for each incoming vehicle we assign a
parameter which determines the vehicle decision to exit along the
incoming direction or leave the roundabout at its first exit by
making a right-turn maneuver. We denote this right-turn
probability by $\sigma_A$ and $\sigma_B$ for incoming A- and
B-vehicles respectively.

\begin{figure}\label{Fig5} \epsfxsize=12.0truecm
\centerline{\epsfbox{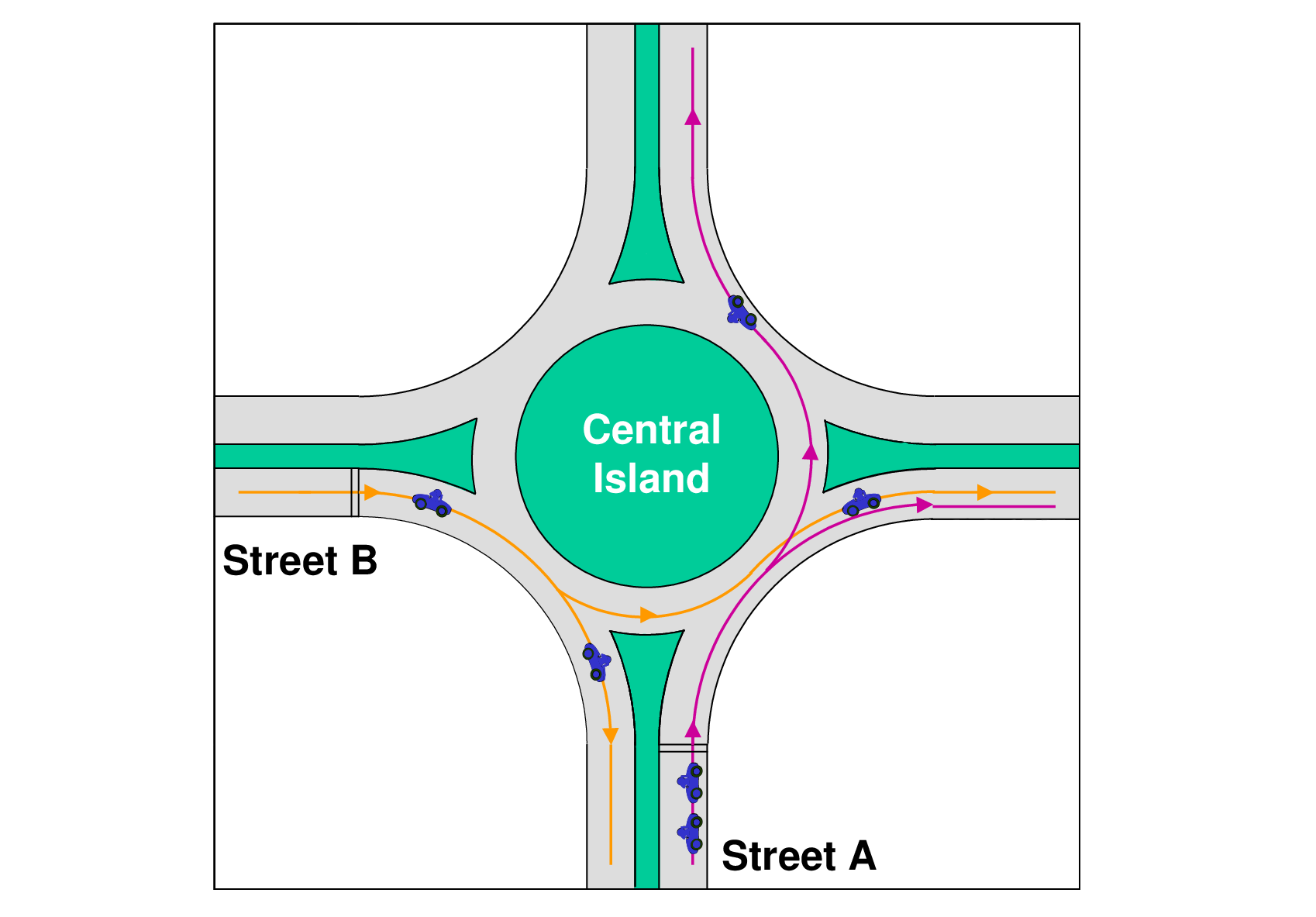}} {\small{Fig.~5: Flow
structure in the roundabout. Right-turn is now permitted and occurs randomly.} }\\
\end{figure}

Before proceeding further, it would be illustrative to discuss the
effect of displaying indicators. By the usage of indicators, each
approaching vehicle can inform the others of his exit direction.
Displaying the right-indicator corresponds to the case in which
the driver intends to make a right-turn and leave the roundabout
at the first exit. Those drivers who intend to exit straight ahead
should not display their indicators. Indicator usage gives rise to
an easier entrance to the roundabout. More specifically, consider
a waiting A-vehicle which is yielding to the B-flow. If this
stopped vehicle observes the displaying indicator of the
approaching B-vehicle, then it, on a deterministic level, knows
that the B-vehicle would not conflict with him (it exits from the
roundabout by the south-bound of street A). Consequently, the
waiting A-vehicle can enter the roundabout simultaneously with the
B-vehicle without any conflict; whereas, without the use of an
indicator, the A-vehicle should wait until the the B-vehicle exits
the roundabout. This would unfavourably increase the waiting
times. The above argument predicts that the usage of an indicator
leads to an easier entrance to the roundabout. Although this
effect locally decreases the waiting times, our simulation
results, nevertheless, prove the contrary. The following graph
shows the waiting time for the symmetric situation in which the
turning probability is taken equally $\frac{1}{2}$ for both A and
B vehicles. Roundabout size is taken 24 cells.

\begin{figure}\label{Fig6}
\epsfxsize=8.5truecm \centerline{\epsfbox{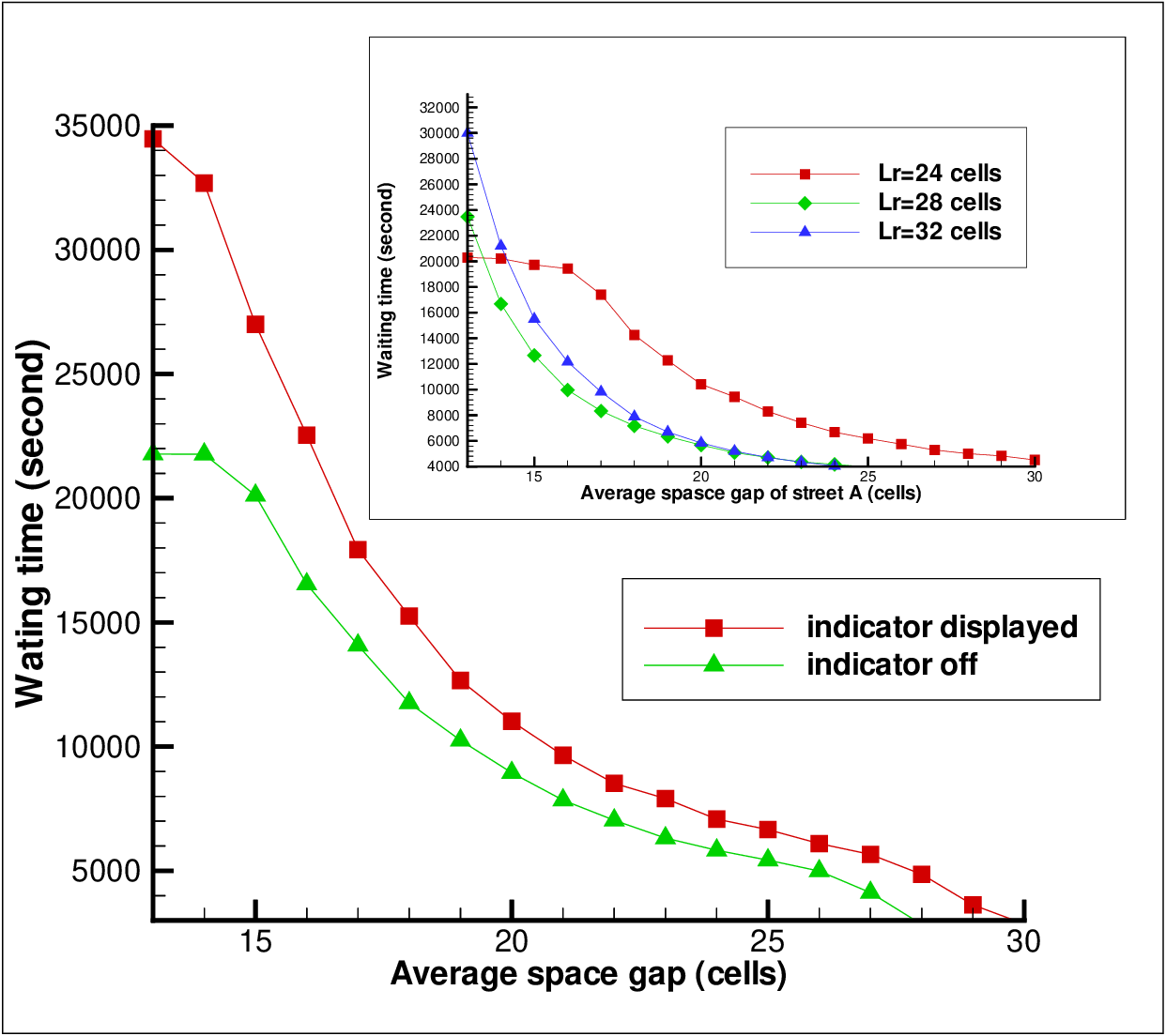}}
\vspace{0.02cm} {\small{Fig.~6: Delay versus average space gap of
approaching cars (symmetric in-flows) for two cases of indicators
displayed and off. The right-turn probability is equal to 0.5 for
approaching cars of both streets. The inset sketches the delay in
terms of $\lambda_A$ for various sizes of the roundabout.
Indicators are off, right-turn probabilities are 0.5 and
$\lambda_B=13$ cells. } }\\
\end{figure}

 In spite of a more convenient entrance to the roundabout by displaying indicators,
the graph above depicts that the impact of displaying indicator
gives rise to increasing the overall delay. This observation can
be explained by the fact that although displaying indicators make
cars enter the central island more conveniently, but this leads to
increase of car density in the central island which
correspondingly increases the probability of blocking the in-flow
direction due to yielding effect. Blocking effect is the dominant
factor and leads to an overall increase of delay. We have also
investigated the dependence of delay on the probability of
right-turns for A-vehicles. The following graph shows the result.

\begin{figure}\label{Fig7}
\epsfxsize=8.5truecm \centerline{\epsfbox{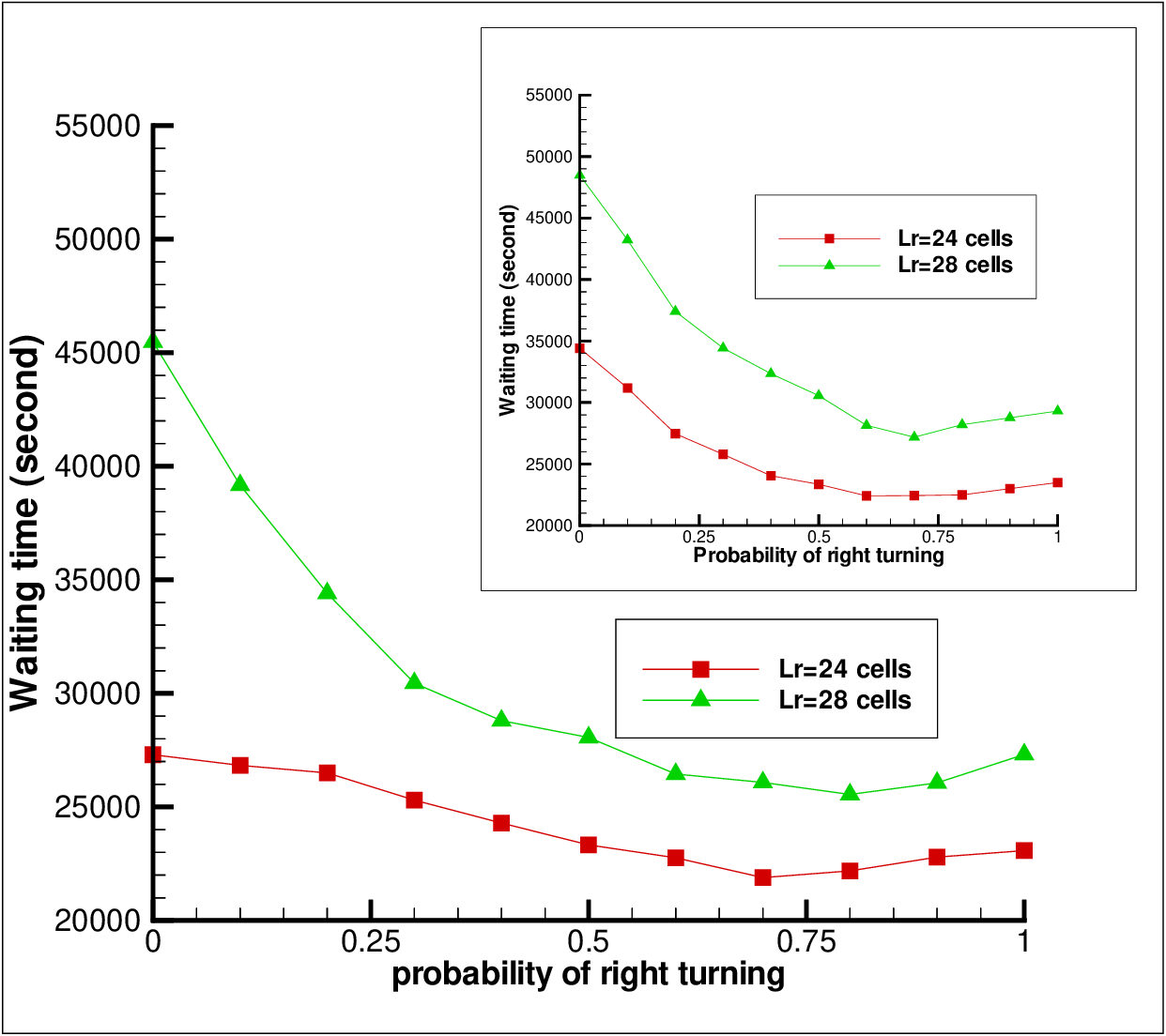}}
\vspace{0.02cm} {\small{Fig.~7: Delay versus right-turn
probability for various roundabout sizes. $\lambda_A=\lambda_B=15$
cells and indicators
are off. In the inset indicators are displayed.} }\\
\end{figure}

According to the results, the delay in minimal for a
size-dependent $\sigma$. It would be useful to discuss the results
of varying the speed limit on the performance of the roundabout.
Evidently, imposing restrictions on the speed limit of vehicles
leads to considerable effects on the waiting time. The lower speed
limit of incoming cars decreases the growth rate of queues which
in turn decreases the waiting times. The following figure
illustrates the situation for various amount of $v_{max}$. Lower
speed limit leads to less delay.

\begin{figure}\label{Fig8}
\epsfxsize=8.0truecm \centerline{\epsfbox{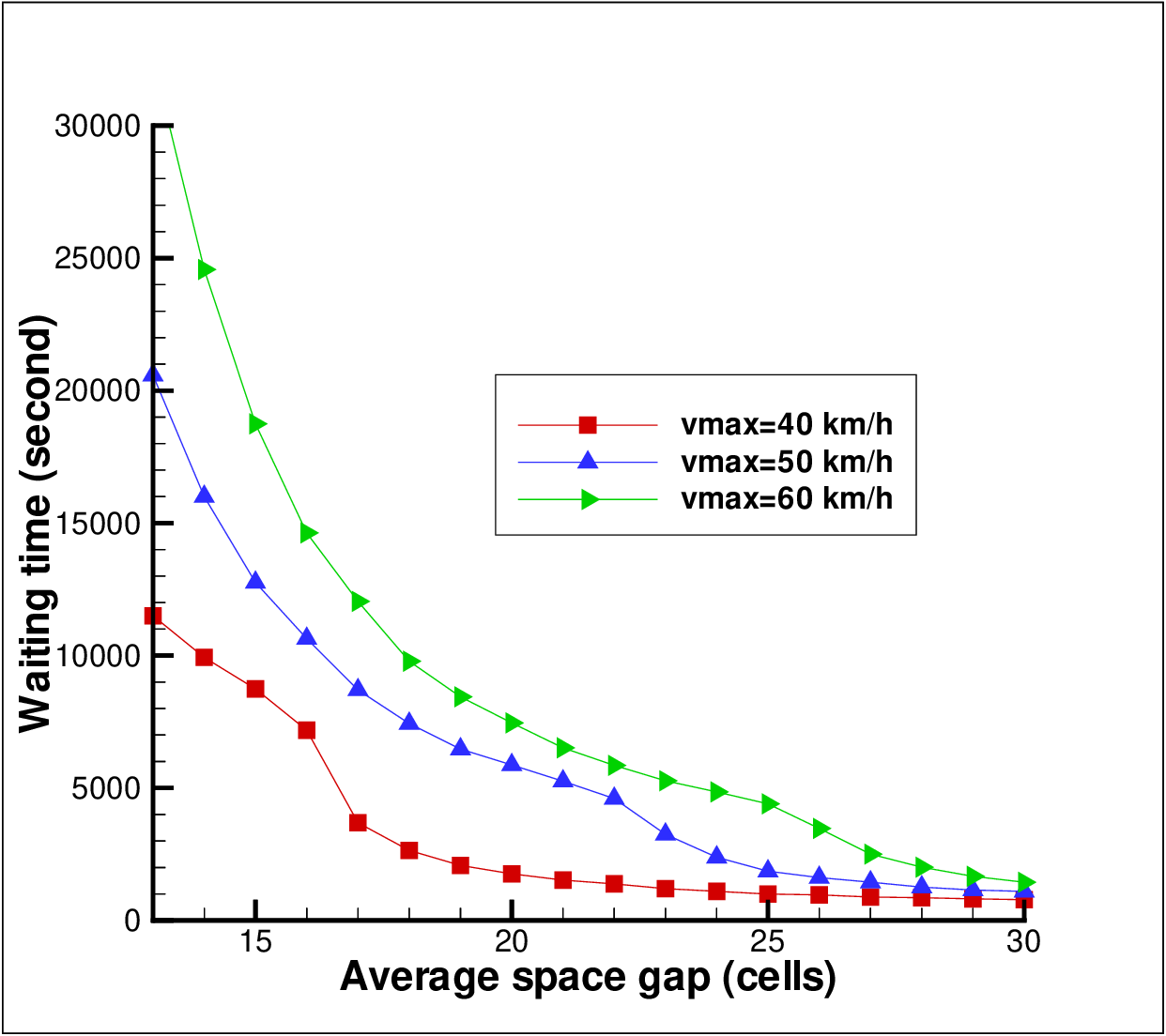}}
\vspace{0.02cm} {\small{Fig.~8: Delay as a function of average
space gap sketched
for various speed limits in the vicinity of the roundabout. Roundabout size is 32 cells.} }\\
\end{figure}

We have also examined the dependence of delay on maximum velocity
for different roundabout sizes. According to the results, one
reaches an asymptotic value independent of the maximum velocity.
The following graph depicts the situation.
$\lambda_A=\lambda_B=15$ cells and right-turn is not permitted.
The results could be of practical results in imposing limitations
on the speed of approaching cars. According to the results, the
lower speed limits yields to improvement of delays. This is
expected since rapid approach to the roundabout causes the cars to
more accumulation which would lead to the increment of delay.

\begin{figure}\label{Fig9}
\epsfxsize=8.5truecm \centerline{\epsfbox{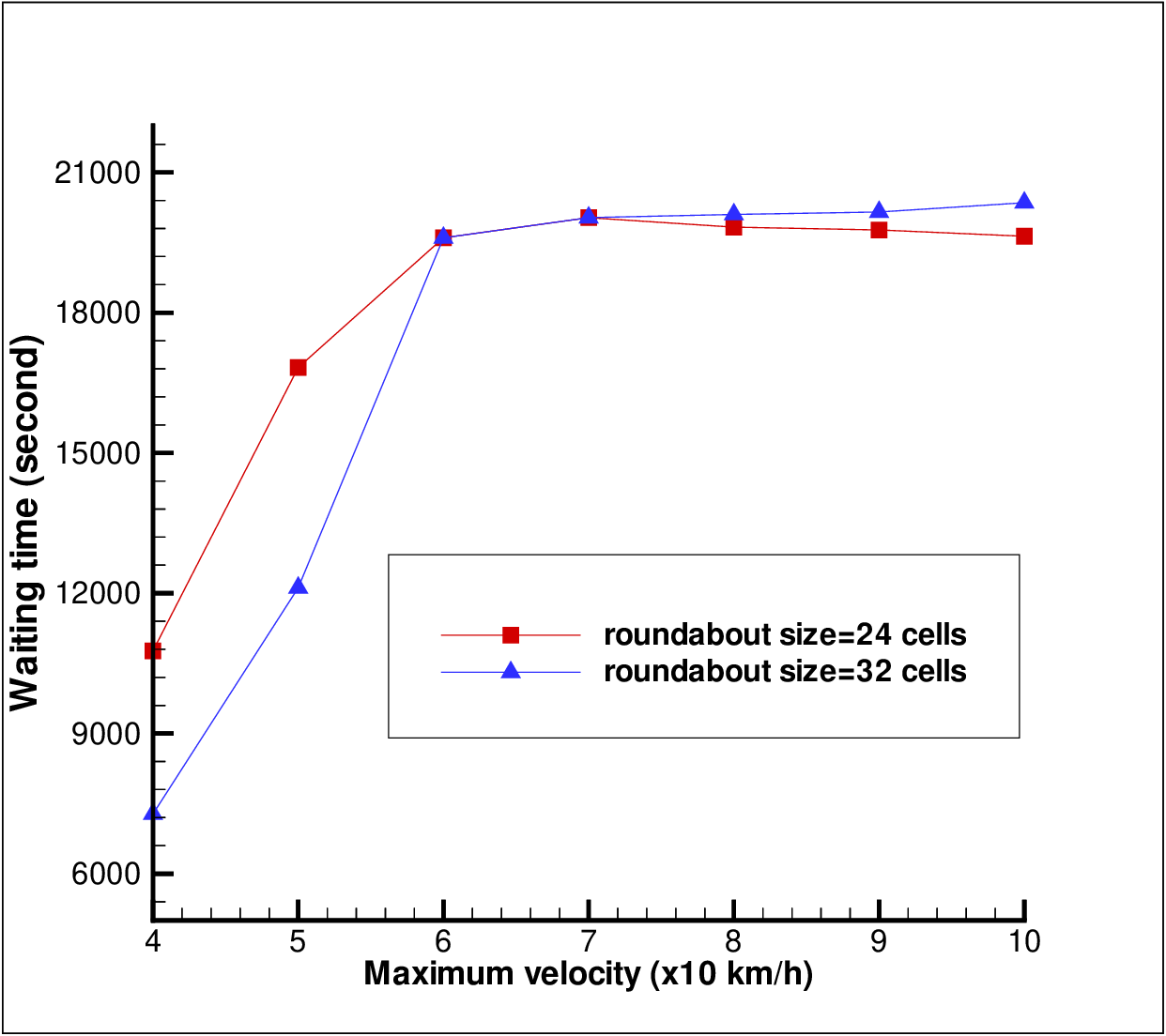}}
\vspace{0.02cm} {\small{Fig.~9: Delay versus maximum velocity for
two values of
roundabout size.} }\\
\end{figure}

\section{Left and U-turn around the island}

Let us now consider a more realistic situation. In its most
general form, vehicles can enter from four directions i.e., north,
south, east and west, to a roundabout. We denote these entries by
$S_{in},N_{in},W_{in}$ and $E_{in}$ respectively. Moreover, there
are four exit directions denoted by $S_{out},N_{out},W_{out}$ and
$E_{out}$. Entering vehicles can exit from any of the outgoing
directions by making an appropriate turning maneuver around the
central island. Let us assume that vehicles enter only from
$S_{in}$ and $W_{in}$ but can exit from every out-going directions
upon their decision (see the following figure). In addition, to
each incoming vehicle we assign a label which determines its exit
direction. These labels are assigned in a random manner. To be
more specific, for each incoming vehicle we let four numbers
$P_S^{\tau}, P_E^{\tau}, P_W^{\tau}$ and $P_N^{\tau}$ denote the
exit probabilities from south, east, west and north exits
respectively. The index $\tau=A,B$ denotes the entrance direction.
We note that for each value of $\tau$ we have $P_S^{\tau}+
P_E^{\tau}+ P_W^{\tau} +P_N^{\tau}=1$.

\begin{figure}\label{Fig10}
\epsfxsize=11.0truecm \centerline{\epsfbox{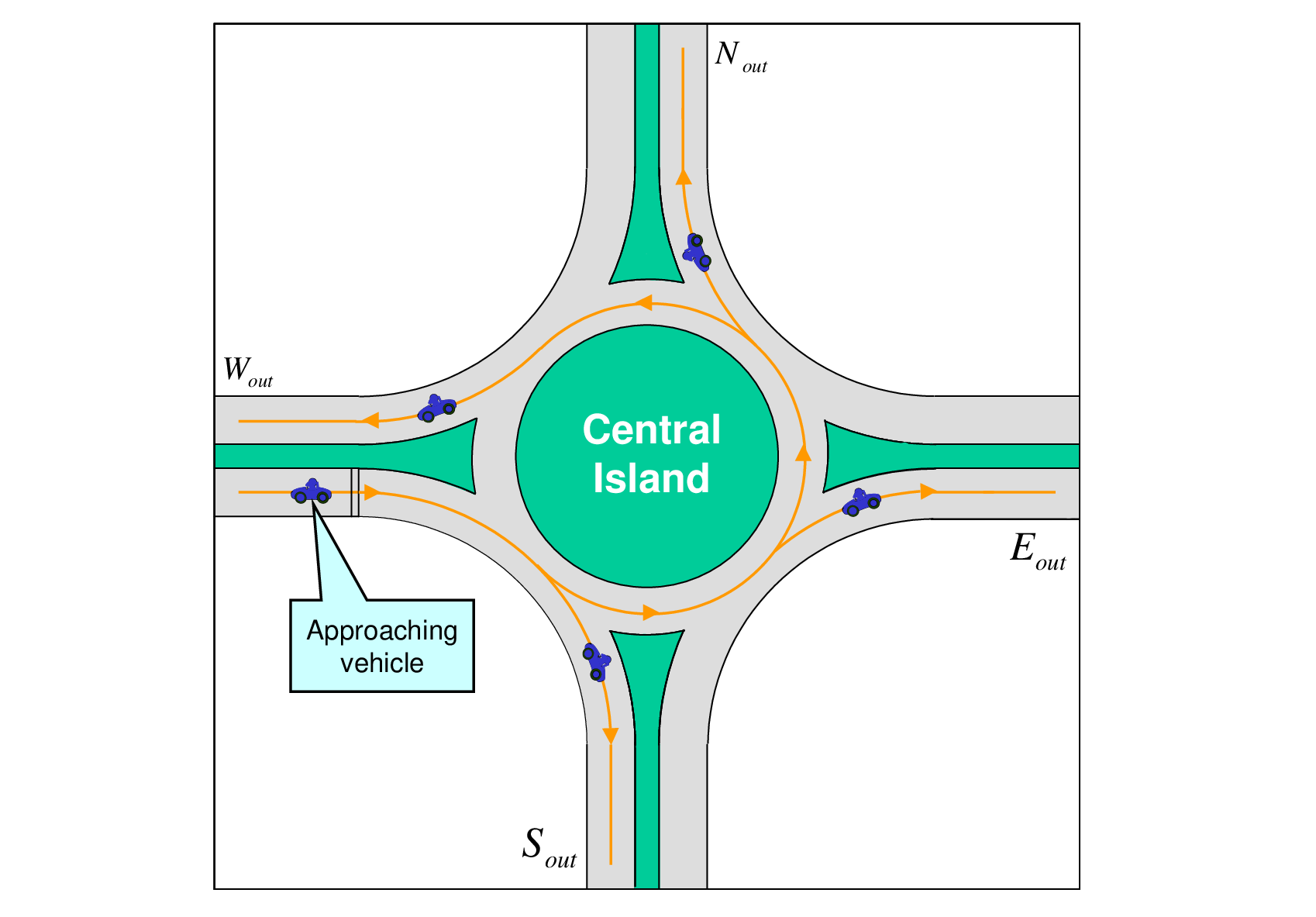}}
\end{figure}

\begin{figure}\label{Fig11}
\epsfxsize=7.5truecm \centerline{\epsfbox{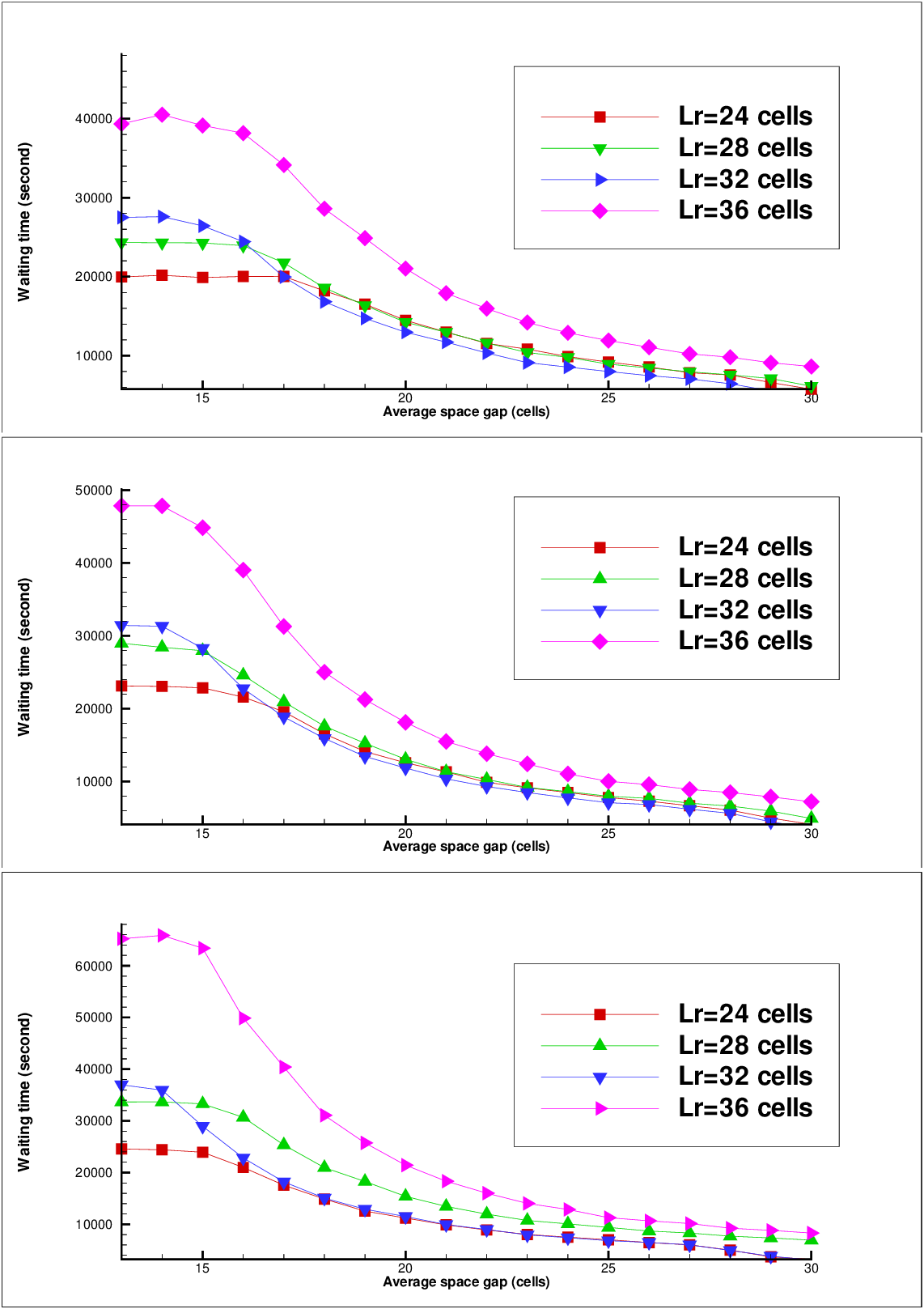}}
\vspace{0.02cm} {\small{Fig.~11: Overall delay versus average
space gap (equal for both streets). Exit probabilities are taken
equal for all the exit directions (top). In the middle graph, the
probability of straightforward exit is 0.5 (preferential exit
direction) and the probability of exit from the remaining exit
directions are equally taken as $\frac{1}{6}$. In the bottom
graph, there are two major exit directions (forward and left) with
probabilities 0.4, 0.4
respectively and two minor ones (backward and right) with probabilities 0.1 and 0.1.} }\\
\end{figure}

In this case, B-vehicles should also yield to traffic in the
roundabout since those vehicles intending to exit from the
$S_{out}$ have priority with respect to incoming cars from
$W_{in}$ entry i.e., B-vehicles. Consequently, in this general
case, both B and A-vehicles contribute to delay. Figure (11)
exhibits the overall delay for a one-hour performance as a
function of average space gap of entering vehicles (taken equal
for both incoming flows) for some choices of roundabout sizes. In
the top graph, exit directions are chosen on an equal basis for
incoming cars $P_S= P_E= P_W= P_N=0.25$ and indicators are assumed
to be off. Maximum velocities are 6 for the outer, and 4 for the
inner vehicles. In the middle and bottom graphs the exit
probabilities are chosen on a biased level. In the following graph
(figure 12) We assume there is a preferential exit direction while
the remaining exit probabilities are the same. Total delay is
sketched for some choices of the preferred exit direction
probabilities. The roundabout circumference is taken to be 24
cells and the in-flows are equal to each other.

\begin{figure}\label{Fig12}
\epsfxsize=8.5truecm \centerline{\epsfbox{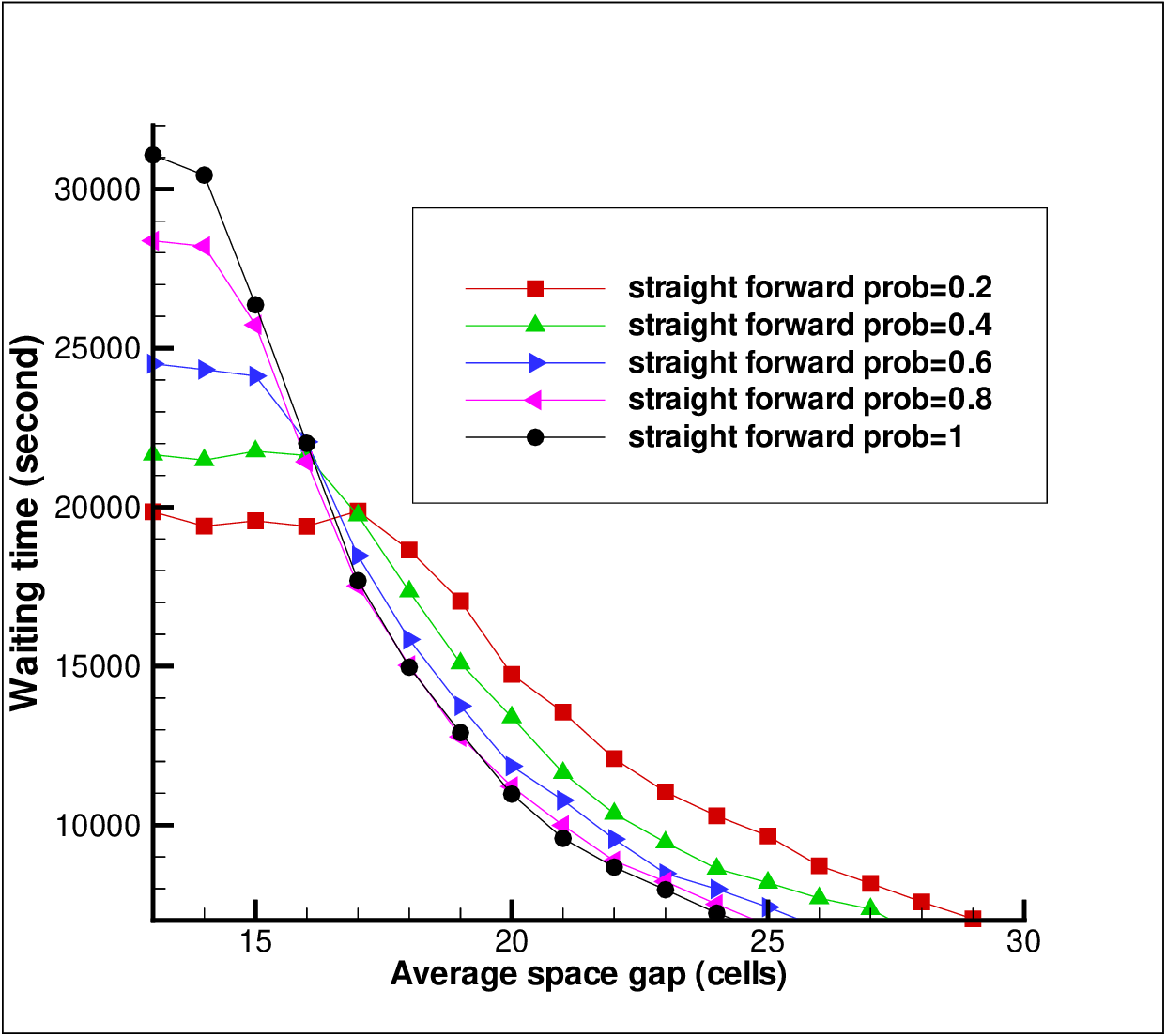}}
\vspace{0.02cm} {\small{Fig.~12: Overall delay in terms of in-flow
for a fixed roundabout size. Probability of forward exiting is
varied. The probability of exit from the remaining directions are
taken to be equal to each other.} }\\
\end{figure}

In the following graph we draw the dependence of one-hour overall
waiting time in terms of the probability of straight exit from the
roundabout. For each approaching vehicle, the probability of
right, left and U-turns are assumed to be equal and the in-flows
are assumed to be symmetric. The top graph corresponds to
$\lambda=13$ cells, in the middle graph, $\lambda$ equals 20 cells
and finally in the bottom graph $\lambda$ is chosen at 28 cells.

\begin{figure}\label{Fig13}
\epsfxsize=8.5truecm \centerline{\epsfbox{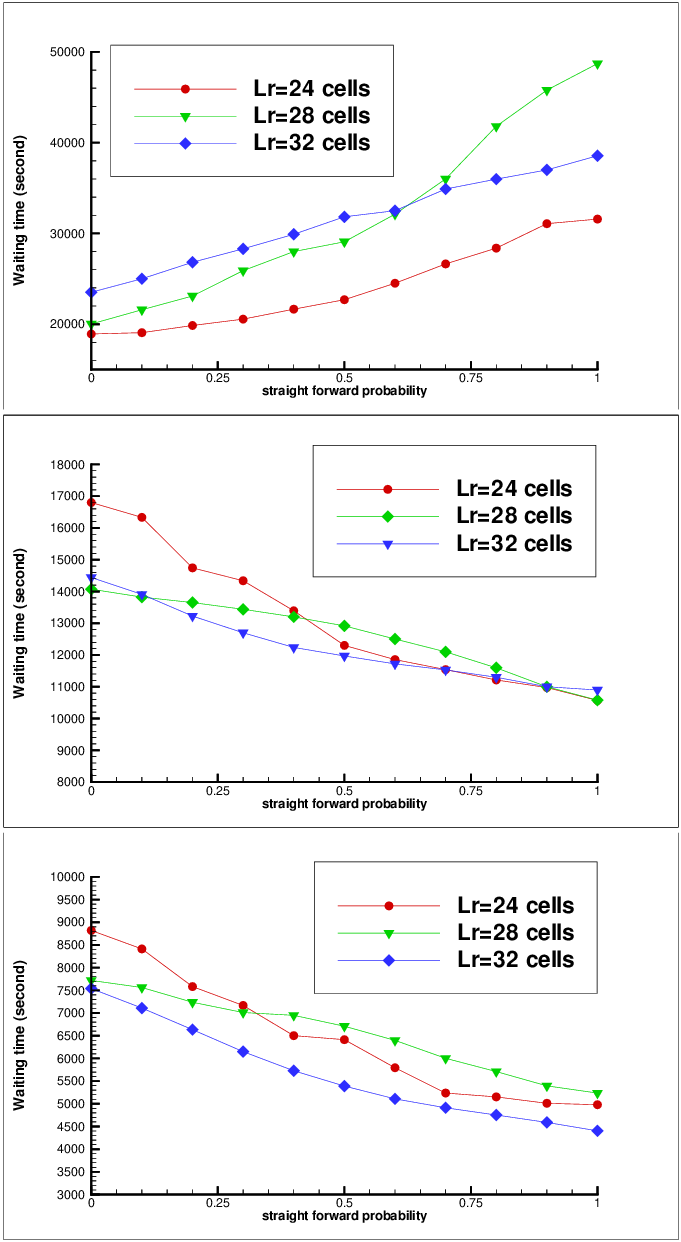}}
\vspace{0.02cm} {\small{Fig.~13: Overall delay in terms of the
forward exit probability for various roundabout sizes. The
in-flows are equal for both streets. Average space gap is
$\lambda=13$ (top), 20
(middle) and 28 (bottom). Exit from the remaining directions occurs on an equal basis.} }\\
\end{figure}

For the sake of comparison, we have simulated four cases
corresponding to different exit situations. We describe the exit
scenarios for A-vehicles. In case one, $P_S=P_W=P_E=0$ and
$P_N=1$. In the second case we have $P_S=P_W=0$ and $P_N=P_E=0.5$.
For the third case $P_S=0$ and $P_E=P_N=P_W=0.33$ and finally the
fourth case considers $P_S=P_W=P_E=P_N=0.25$. Similar arguments
apply to B-vehicles. The following figure depicts the delay curve
for these four exit cases as a function of the inverse traffic
in-flow rate. The results show that right-turn exit is the main
factor in reducing total delay which is justified by the its least
conflict.

\begin{figure}\label{Fig14}
\epsfxsize= 8.5truecm \centerline{\epsfbox{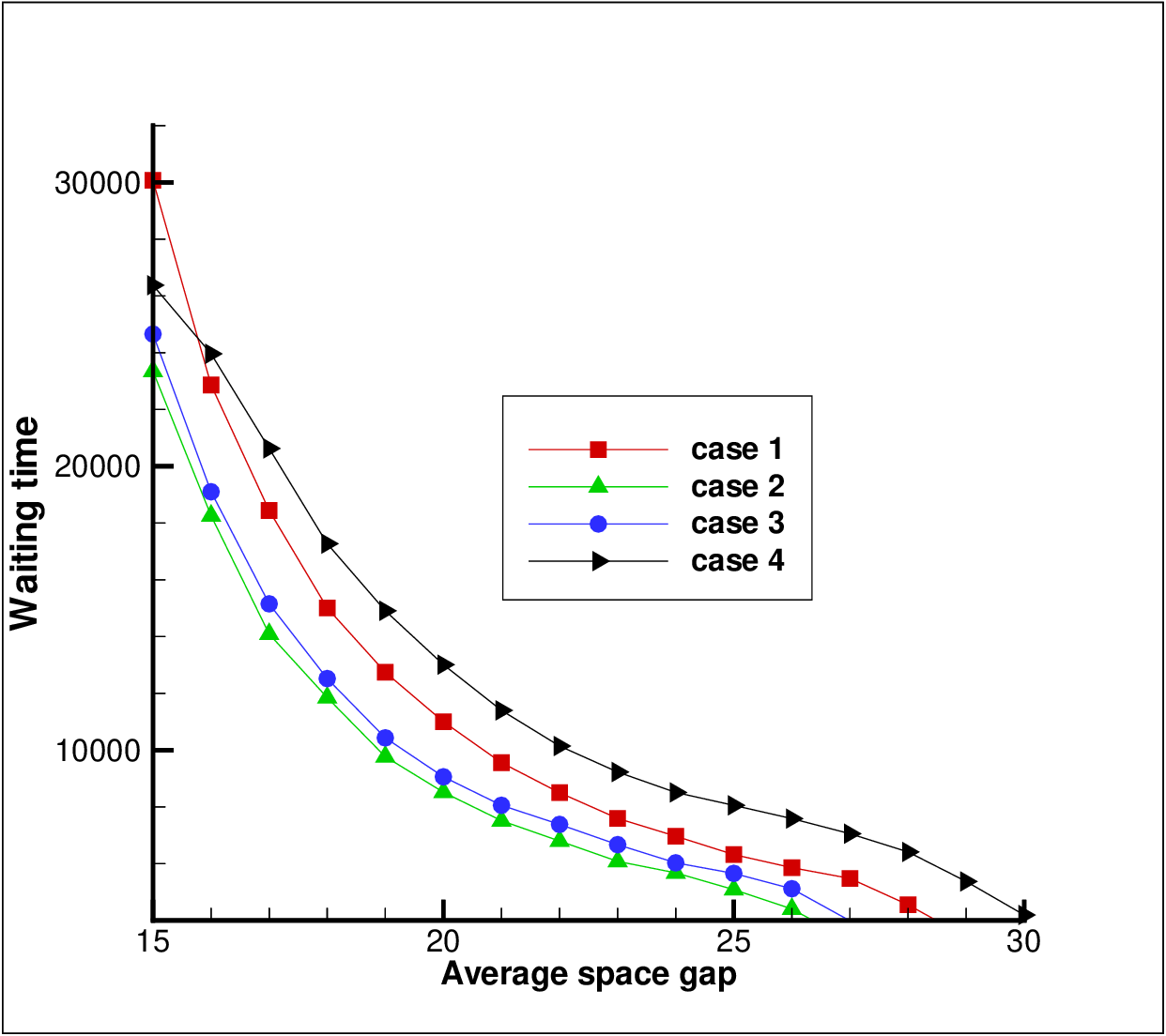}}
\vspace{0.02cm} {\small{Fig.~14 : Overall delay versus equal
average space gap of in-flows. Exit probabilities are equal in
each four cases.} }\\
\end{figure}

\section{ Comparison to other controlling schemes }

Let us now compare the roundabout performance with signalised
control methods of an intersection. This comparison is our main
motive for studying roundabout characteristics. Let us replace the
roundabout with an intersection with traffic lights. For
simplicity, we consider the intersection of two one-way to one-way
streets which are assumed to direct single-lane traffic flow.
Basically there are two types of signalisation: {\it fixed-time}
and {\it traffic adaptive}. We first describe the fixed-time
method. In this control scheme, the traffic flow is controlled by
a set of traffic lights which are operated on a fixed-cycle. The
lights periodically turn green with a fixed period (cycle length)
$T$. This period is divided into two parts: in the first part, the
traffic light is green for street $A$ (simultaneously red for
street $B$). This part lasts for $T_{g}$ seconds ( $ T_{g} < T $).
In the second part, the lights change colour and movement is
allowed for the cars of road B. The second part lasts from $T_{g}$
to $T$. This behaviour is repeated periodically. In
\cite{foul-sadj-shab} and \cite{foul-nemat} we have shown that the
optimal traffic flow is obtained at:
$T_g^{opt}=\frac{\alpha_A}{\alpha_A+\alpha_B}T $
 where $\alpha_A$ and $\alpha_B$ denote traffic volumes in street A and B respectively.
This implies that the optimal green time given to street A should
be proportional to its in-flow rate. In figure (16) we compare the
performance of the corresponding roundabout with fixed-time
signalisation strategy. Traffic volumes are assumed to be equal
for both streets. Furthermore, we assume that incoming vehicles
cannot turn and should move straight ahead.

\begin{figure}\label{Fig15}
\epsfxsize=8.2truecm \centerline{\epsfbox{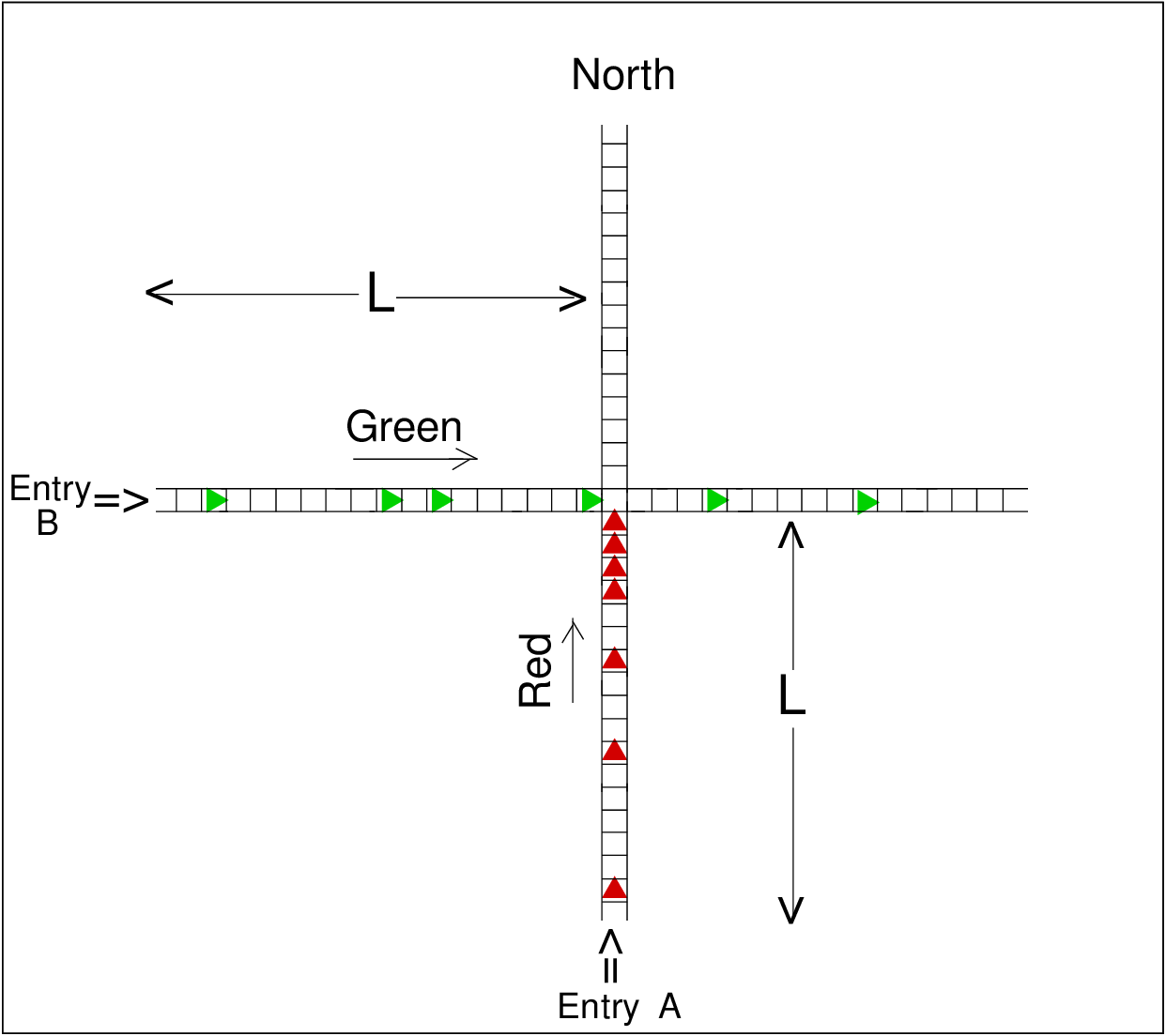}}
\vspace{0.02cm} {\small{Fig.~15: One-way to one-way
intersection with traffic lights.} }\\
\end{figure}

\begin{figure}\label{Fig16}
\epsfxsize=8.5truecm \centerline{\epsfbox{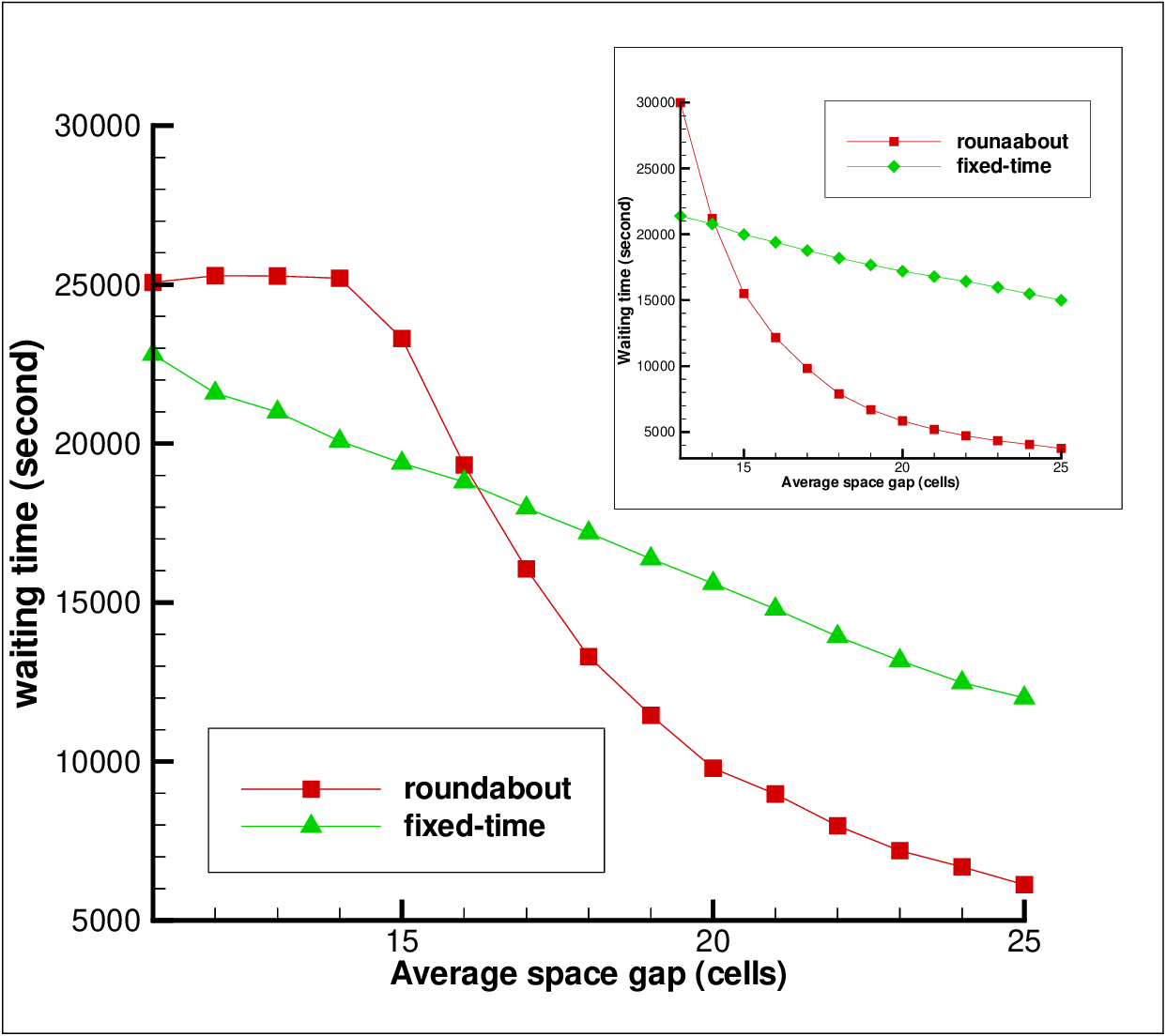}}
\vspace{0.02cm} {\small{Fig.~16: In signalised controlling the
period of lights are 40 seconds and the green times are equally
distributed to the streets. The inset describes an asymmetric
situation in the in-flows. Average space gap of approaching cars
are fixed at $\lambda_A=13$ cells while $\lambda_B$ is varied. Roundabout size is 24 cells.} }\\
\end{figure}

According to the above graph, in relatively light traffic states,
characterized by a large average space gap, a roundabout shows a
better performance and gives rise to lower delays. Conversely, in
more congested traffic situations, controlling the intersection by
signalised traffic light leads to better results. Our simulation
results give the critical in-flow rate below which the
intersection should be controlled in a self-organized manner. This
result can be explained by noting that in sufficiently light
traffic states, the approaching cars can easily find the required
space gap in the flow of conflicting direction hence they can
enter the roundabout without spending much times whereas in a
signalised scheme, they have to wait at the red parts of the
signal even if the flow is negligible in the conflicting
direction. This proves that below a certain congestion, the
roundabout efficiency is higher than fixed-time signalised. We now
discuss the traffic adaptive controlling scheme in which the light
signalisation is adapted to the traffic at the intersection.
Nowadays, advanced traffic control systems anticipate the traffic
approaching intersections. These adaptive systems have the
capability to dynamically modify the signal timing in response to
fluctuating traffic demand. Traffic-responsive methods have shown
a very good performance in controlling city traffic, and now a
variety of schemes exists in the literature
\cite{robertson,porche,huberman}. In these schemes, the data
obtained via traffic detectors installed at the intersection is
gathered for each movement direction, and it is possible to count
the queue-lengths formed behind the red lights. One can also
measure the time-headways between successive cars passing each
lane; thus, it is possible to estimate the traffic volume existing
at the intersection. There are various methods for the
distribution of green times. In what follows we try to explain
some standard ones. In each scheme, the green time of a typical
direction is
terminated if some conditions are fulfilled:\\

{\bf Scheme (1)}: The queue length in the conflicting direction
exceeds a cut-off value $L_c$. This scheme only adapts to
the traffic states on the red street.\\

{\bf Scheme (2)}: The global car density on the green street falls
below the cut-off value $\rho_c$. Here the algorithm
solely adapts to the traffic state in the green street.\\

{\bf Scheme (3)} : Each direction is endowed with two control
parameters $L_c$ and $\rho_c$. The green phase is terminated if
the conditions: $ \rho ^{g} \leq \rho_c $ and $L^{r} \geq
L_c$ are both satisfied.\\

In scheme three the algorithm implements the traffic states in
both streets. The superscripts "r" and "g" refer to words "red"
and "green" respectively. We note that the first two schemes are
special cases of the more general scheme (3). Schemes (1) and (2)
are the limiting cases of schemes (3) by letting $\rho_c
\rightarrow 1$ and $L_c \rightarrow 0$ respectively. In general,
the numerical value of control parameters $L_c$ and $\rho_c$ could
be taken different for each individual street. In what follows we
present our simulation results for some types of adaptive
signalisation schemes introduced above, and compare them to a
self-organized scheme by roundabout.

\begin{figure}\label{Fig17}
\epsfxsize=8.5truecm \centerline{\epsfbox{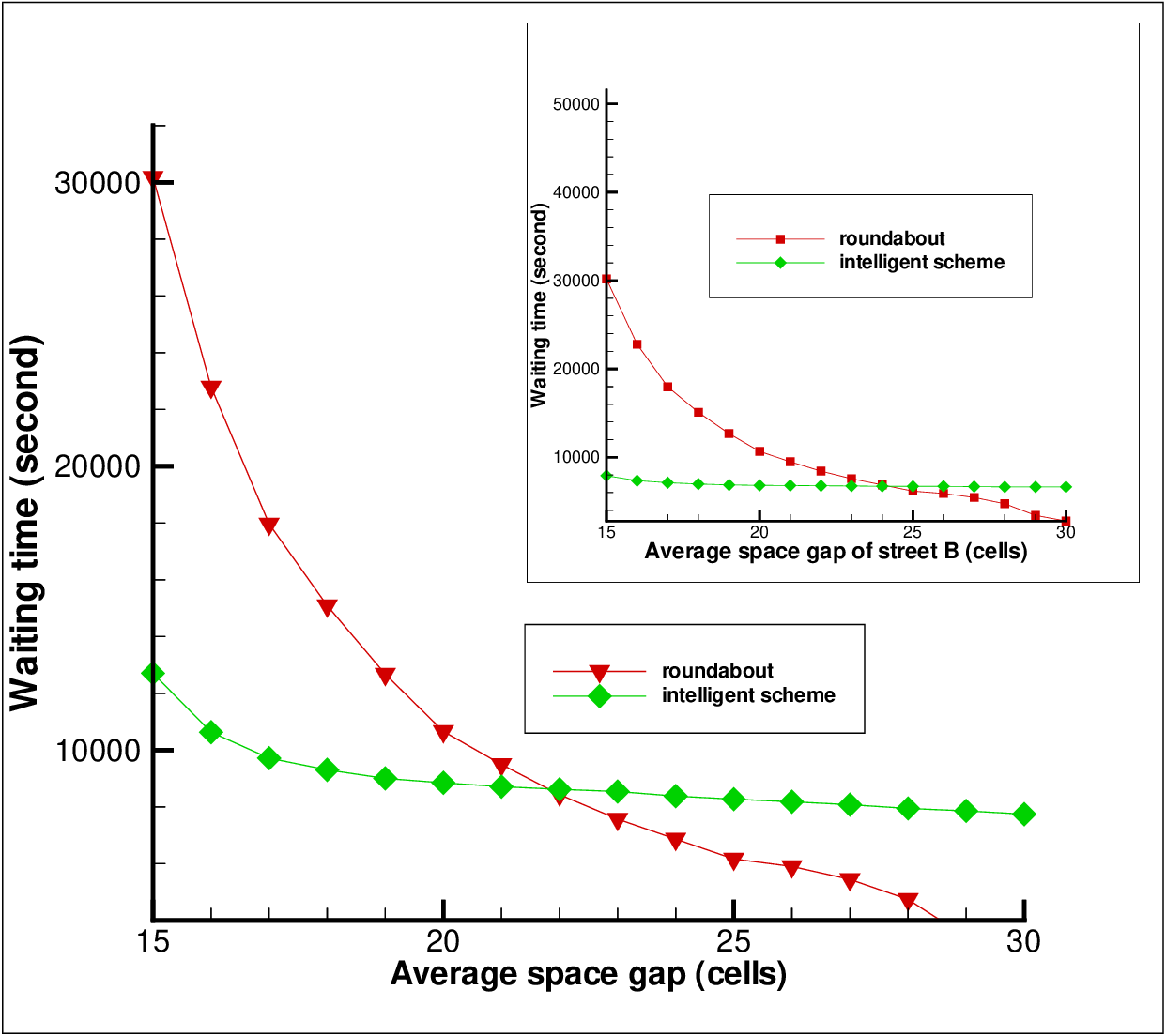}}
\vspace{0.02cm} {\small{Fig.~17: Overall delay in terms of
symmetric average space gap of approaching vehicles. Roundabout
size is 24 cells. The critical cut-off length in adaptive strategy
is $L_c=5$. The inset describes an asymmetric situation in the
in-flows. Average space gap of approaching cars are fixed at
$\lambda_A=13$ cells while $\lambda_B$ is varied.} }\\
\end{figure}

Analogous to fixed-time method, here we see that below a certain
traffic volume, roundabout is more efficient. We note that in the
adaptive scheme, the numerical value of critical $\lambda$ is
considerably reduced with respect to fixed-time method. This is
due to the advantage of adaptive schemes over fixed-time ones.
This comparison has thoroughly been discussed in
\cite{foul-sadj-shab}. Fixed-time predicts $\lambda_c=16$ cells
while in adaptive method it goes to  $\lambda_c=22$.
\section{Summary and Concluding Remarks}

Traffic signal control is a central issue in the design of
advanced traffic management systems. Recent strides in the
modelling and in the simulation of traffic flow have opened new
possibilities for traffic control and optimization. In this
regard, the micro-simulation of city traffic could be of practical
relevance for various applications in urban traffic. Isolated
intersections are fundamental operating units of the sophisticated
and correlated urban network, and their thorough analysis would be
advantageous towards the ultimate task of the global optimization
of the city network. While signalised intersections are
traditional objects designed for controlling the traffic flow in
conflicting directions, modern roundabouts, which have recently
been designed and operated, account for an alternative strategy
for traffic control. Enthusiasm for safety and for the high
capacity of roundabouts has resulted in a huge increase in the
number of roundabouts. In contrast, as growing traffic demand
causes non-conforming traffic circles to fail, they are converted
to other types of intersections. In addition to the features that
characterize a modern roundabout, yield-at-entry, deflection and
flare, roundabouts often have other important safety features.
Although some people involved in vehicular traffic may be
uncomfortable initially with the idea of a traffic roundabout, it
is a solution that is environmentally friendly and requires less
in annual maintenance costs since it replaces traffic signals.
Nevertheless, the efficiency of roundabouts is still under debate,
and many experts believe that signalised intersections show a
better performance in most circumstances. To settle this debate,
at least to a partial degree, we have tried to quantitatively
explore the basic features of roundabout in order to have a better
insight into the problem. In this paper we have investigated the
characteristics of traffic at an isolated roundabout in the
framework of cellular automata. For this purpose, we have
developed and analyzed the performance of the various aspects of
roundabouts, the most important of which is delay. Our simulation
shows that overall delay is significantly affected by roundabout
size. Our simulations gives the optimal size for various traffic
volumes. Another relevant aspect is the {\it indicator displaying
}, which noticeably affects the overall delay. The major
conclusion made from our simulation results proves the existence
of critical congestion, dominated by the statistics of arrival
space gaps, over which the intersection is made more efficient by
signalisation strategies. In a more realistic situation, the flow
can circulate around the central island via an additional lane.
The interior lane should be used by those vehicle intending to
make left or U-turns while the exterior one should be taken by
those drivers who tend to turn right or move straightforward. The
second interior lane may drastically change the behaviour of
indicator displaying thus leading to improvement of delay. In the
present simple case of single-lane circulation, our simulations
implies that injection of vehicles from more than two entries
leads to global blocking of flows and growing delays. This effect
is due to the saturation of circulating flow which hinders the
incoming fluxes. Implementation of additional interior lane will
certainly removes the blocking and gives rise to realistic
results. In this general situation, roundabout performance
undergoes fundamental changes, and many interesting phenomena
arise which we are still currently exploring.

\section{ Acknowledgments}

 We would like to express our gratitude to Gunter M. Sch\"{u}tz for bringing this problem to
 our attention and for his fruitful and enlightening discussions. We wish to thank
 Shahrokh Zamayeri for his valuable helps. Special
 thanks go to Richard W. Sorfleet for reading the manuscript.

\bibliographystyle{unsrt}

\begin{thebibliography}{99}


\bibitem{css99}
D. Chowdhury, L. Santen and A. Schadschneider, {\em Physics
Reports}, {\bf 329}, 199 (2000).


\bibitem{helbbook} D.\ Helbing,  {\em Rev. Mod.
Phys.}, {\bf 73}, 1067; {\em Vehrkersdynamik: Neue Physikalische
Modellierungskonzepte}, Springer, Berlin, 1997.


\bibitem{kerner} B.S. Kerner, {\em Networks and Spacial Economics}, {\bf 1},
35 (2001).


\bibitem{book}
 {\em Traffic Flow Fundamentals}, Prentice Hall (1990) by A.D.\
May. {\em Transportation and Traffic Theory}, Elsevier (1993) by
C.F \ Daganzo.


\bibitem{tgf95} D.E.\ Wolf, M.\ Schreckenberg and A.\ Bachem (eds) {\em
Traffic and granular flow} (World Scientific, Singapore, 1996).



\bibitem{tgf97} D.E\ Wolf and M.\ Schreckenberg (eds.) {\em Traffic and
granular flow} (Springer, Singapore, 1998).



\bibitem{tgf99} \ H.J.\ Herrmann, D.\ Helbing, M.\
Schreckenberg and D.E.\ Wolf (eds.) {\em Traffic and Granular
flow}
 (Springer, Berlin, 2000).






\bibitem{robertson} D.I. Robertson and R.D. Bretherton, {\it Optimizing
networks of traffic signals in real-time: the SCOOT method }, {\em
IEEE Transportation on Vehicular Technology}, {\bf 40}, 11 (1991).



\bibitem{porche}I. Porche, M. Sampath, Y.-L. Chen, R. Sengupta and S.
Lafortune, {\em A decentralized scheme for real-time optimization
of traffic signal } in the proceeding of 1996 {\it IEEE
International Conference on Control Applications}, 582-589.


\bibitem{huberman} B. Faieta and B.A. Huberman, {\it firefly : A
synchronisation strategy  for urban traffic control}, Xerox Palo
Alto Research Center, Palo Alto, CA 94304.



\bibitem{knospe} W. Knospe, L. Santen A.\ Schadschneider and
and M.\ Schreckenberg, {\em J. Phys. A: Math. Gen}, {\bf 33},
L477.





\bibitem{ns92} K. Nagel, M. Schreckenberg, {\em J.Phys.I France} {\bf 2},
2221 (1992).



\bibitem{erlang} S. Kotz and N.L Johonson, {\em Encyclopedia of
Statistic Sciences}, vol 3, 292 (1983).


\bibitem{foolad} M.E. Fouladvand and M. Nematollahi, {\em Eur. Phys. J. B}
{\bf 22}, 395 (2001).


\bibitem{foul-sadj-shab} M.E. Fouladvand, Z. Sadjadi and M.R.
Shaebani, to appear in {\em J. Phys. A: Math. Gen}.


\bibitem{scats1} P. Lowrie, {\it SCATS: A Traffic Responsive Method for
Controlling Urban Traffic}, tech. rep., {\em Road and Traffic
Authority}, NSW, Australia.


\bibitem{scats2} A.G. Sims, "SCATS: the Sydney co-ordinated adaptive
system", in the proceeding of the {\it Engineering Foundation
Conference on Research Priorities in Computer Control of Urban
Traffic Systems}, 12, 1979.



\end{thebibliography}

\end{document}